\documentclass{article}
\usepackage{ijcai22}

\usepackage{times}
\usepackage{soul}
\usepackage{url}
\usepackage[hidelinks]{hyperref}
\usepackage[utf8]{inputenc}
\usepackage[small]{caption}
\usepackage{graphicx}
\usepackage{amsmath}
\usepackage{amsthm}
\usepackage{booktabs}
\usepackage{algorithm}
\usepackage{algorithmic}
\usepackage{color}
\usepackage{amsfonts}
\usepackage{bm}
\usepackage{multirow}
\usepackage{float}
\usepackage{subfig}

\title{Transition Relation Aware Self-Attention for Session-based Recommendation}
\author{
Guanghui Zhu$^1$\and
Haojun Hou$^1$\and
Jingfan Chen$^1$\and
Chunfeng Yuan$^1$\And
Yihua Huang$^1$\footnote{Contact Author}
\affiliations
$^1$Nanjing University\\
\emails
\{zgh, cfyuan, yhuang\}@nju.edu.cn,
\{hanjunhou, jingfan.chen\}@smail.nju.edu.cn
}
\date{December 2021}

\begin{document}

\maketitle

\begin{abstract}
    Session-based recommendation is a challenging problem in the real-world scenes, e.g., e-commerce, short video platforms, and music platforms, which aims to predict the next click action based on the anonymous session. Recently, graph neural networks (GNNs) have emerged as the state-of-the-art methods for session-based recommendation. However, we find that there exist two limitations in these methods. One is the item transition relations are not fully exploited since the relations are not explicitly modeled. Another is the long-range dependencies between items cannot be captured effectively due to the limitation of GNNs. To solve the above problems, we propose a novel approach for session-based recommendation, called \textbf{T}ransition \textbf{R}elation \textbf{A}ware \textbf{S}elf-\textbf{A}ttention (TRASA). Specifically, TRASA first converts the session to a graph and then encodes the shortest path between items through the gated recurrent unit as their transition relation. Then, to capture the long-range dependencies, TRASA utilizes the self-attention mechanism to build the direct connection between any two items without going through intermediate ones. Also, the transition relations are incorporated explicitly when computing the attention scores. Extensive experiments on three real-word datasets demonstrate that TRASA outperforms the existing state-of-the-art methods consistently.
\end{abstract}
\section{Introduction}
\begin{figure}
    \centering
    \includegraphics[width=0.9\linewidth]{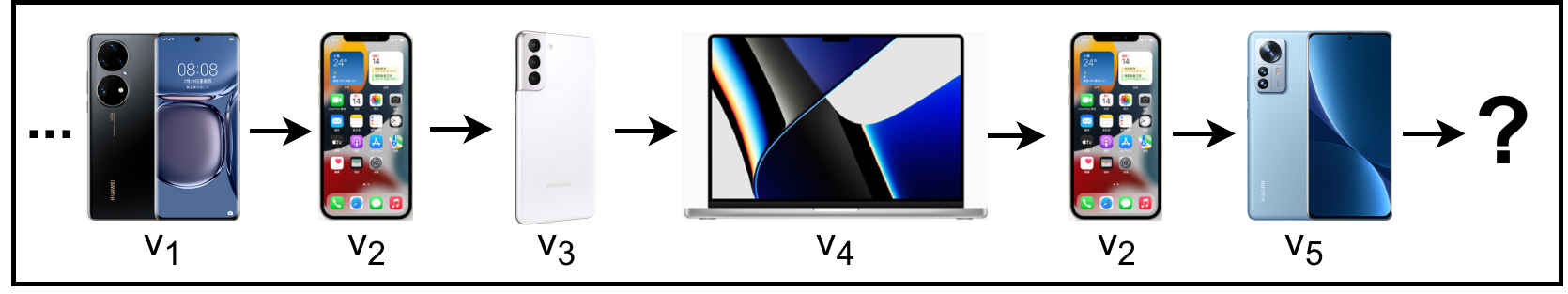}
    \caption{A session when buying a phone}
    \label{fig:session}
    \vspace{-2ex}
\end{figure}

Recommender systems (RS) play a very important role in many real-world web applications, e.g., e-commerce, short video platforms, and music platforms. The reason behind the great success of RS is that it can address the information overload problem by making personalized recommendation for every user. Traditional RS make recommendations based on users' profiles and their all historical activities. However, this information cannot be accessed in some scenarios, where only the behaviors of an anonymous user in a short period is available. To improve the quality of recommendation for anonymous users, session-based recommendation is proposed, which aims to make a better prediction for the next click action based on an anonymous session. Figure \ref{fig:session} shows a session composed of multiple user-item interactions that happen together in a continuous period of time.

Due to its highly practical value, there have been many studies about session-based recommendation~\cite{survey}. Early methods are mainly based on item similarities~\cite{Item-KNN} or Markov Chains~\cite{mdp,fpmc,hrm}. Due to the success of deep learning, many deep learning-based methods have been applied in session-based recommendation. They capture the user's interest by using recurrent neural networks (RNNs)~\cite{gru4rec}, applying attention mechanism~\cite{stamp} or utilizing both~\cite{narm}. Recently, graph neural network (GNN)-based methods have become the most popular methods used in session-based recommendation and have become the state-of-the-arts~\cite{sr-gnn,fgnn,gc-san,lessr,gce-gnn,dat-mdi}. Although the GNN-based methods have achieved promising results in session-based recommendation, we find that they still have two limitations. 

First, the transition relations between items are not fully exploited. As  shown in Figure \ref{fig:session}, $v_5$ is the third item clicked after $v_3$. The transition relation between $v_3$ and $v_5$ is $v_3 \xrightarrow{next} v_4 \xrightarrow{next} v_2 \xrightarrow{next} v_5$. In GNN-based methods, although these transition relations can be reflected in the graph topology to some extent, they are not modeled explicitly in the message passing process.
The position embeddings can capture partial information of transition relations~\cite{gce-gnn}. However, position embeddings treat the session as a sequence and every position is encoded as a unique vector. But the same item may appear multiple times in a session, e.g., $v_2$ in Figure \ref{fig:session}. Thus, it is inappropriate that the same item has multiple different position embeddings in a session.

Second, the long-range dependencies between items can not be captured effectively. Even though GNNs can model graph-structured data, they can not be stacked many layers due to the over-smoothing or over squashing problems~\cite{over-smooth,over-squash}. GNN-based methods usually achieve the best performance using 1 to 3 layers, which means the item dependencies over 3-hop neighbors are hardly captured. However, the length of real-world sessions is usually greater than 3~\cite{survey} and the long-range dependencies do exist. In Figure \ref{fig:session}, even though the distance between $v_1$ and $v_5$ is 5, they still have a strong dependency since they are both phones.
Since a session depicts a user's short-term interest, it is reasonable to assume that every item appeared in the same session should have a strong or weak relationship that cannot be ignored.

To address the two limitations, we propose a novel method for session-based recommendation, called \textbf{T}ransition \textbf{R}elation \textbf{A}ware \textbf{S}elf-\textbf{A}ttention (TRASA). A session sequence is first converted to a graph where the nodes represent different items and edges represent their transition order in the original session sequence. To model the transition relations between items accurately, a relation encoder is introduced to encode all the relations from a graph perspective explicitly. Specifically, it encodes the shortest path between two items by GRU (Gated Recurrent Unit~\cite{gru}) as their transition relation. To capture the long-range dependencies between items, a self-attention mechanism is applied to get the potential relation between any two items without going through intermediate ones. The transition relations are utilized when computing the attention scores. Finally, the graph is reverted to its original sequence to get the final session representation and make prediction. We summarize our contributions as follows:

\begin{itemize}
    \item To the best of our knowledge, we are the first to propose to encode the shortest path between any two items as their transition relations from a graph perspective in session-based recommendation.
    \item We apply a self-attention mechanism to make direct connections between any two items in a session, which can capture the long-range dependencies effectively. And we incorporate the transition relations when calculating the attention scores.
    \item We conduct extensive experiments on three real-world datasets and the results demonstrate that our method outperforms the state-of-the-art methods.
\end{itemize}

\begin{figure*}
    \centering
    \includegraphics[width=\linewidth]{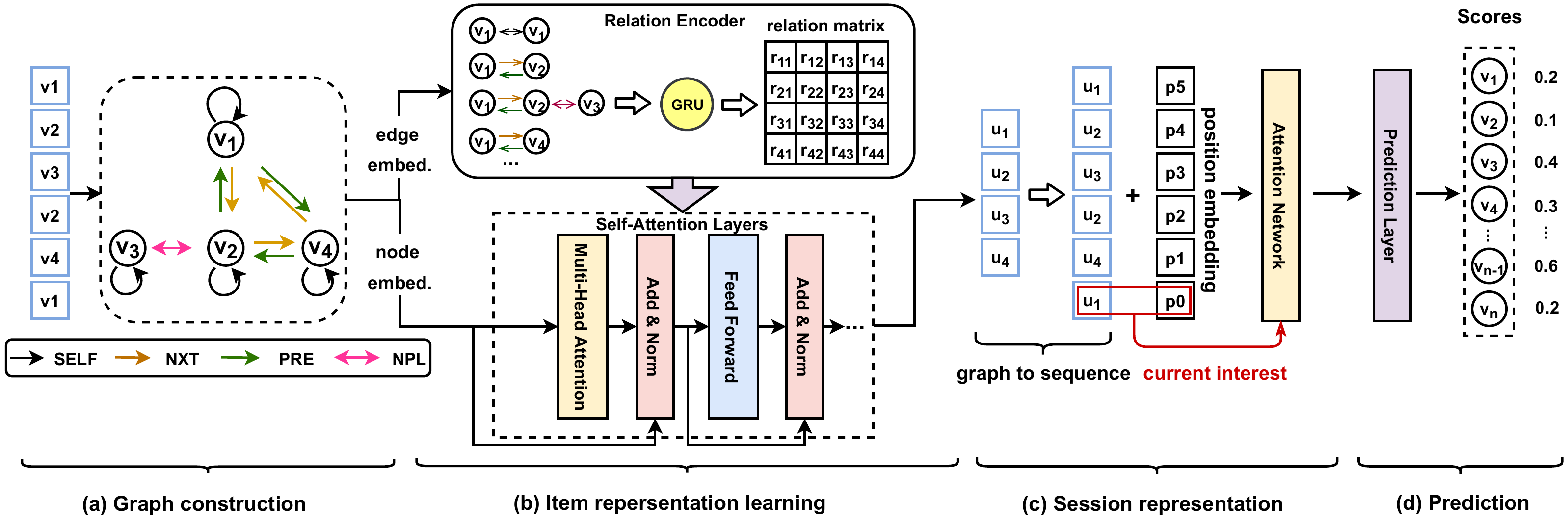}
    \caption{The workflow of the proposed method TRASA. A session is first converted to a graph. Then, the relation encoder encodes the shortest path between items as their transition relations and the self-attention layers utilize these relations to learn item representations. Next, the graph is reverted to is original sequence and position embeddings are incorporated to get the final session representation. Finally, the prediction layer computes the probability for every candidate item.}
    \label{fig:mode_architecture}
\end{figure*}
\section{Related Work}
The existing session-based recommendation methods can be divided into the following categories.

\paragraph{Traditional Methods.}
Since a session is a sequence of items clicked in chronological order, Markov Chain-based methods have been proposed in session-based recommendation. Markov Chain-based methods map a session into a Markov Chain and predict the next clicked item based on the last one. MDP~\cite{mdp} applies Markov decision processes to model the recommendation process as a sequential process. As an improvement, FPMC~\cite{fpmc} utilizes both Markov chains and matrix factorization to capture sequential effects and long-term user taste simultaneously.

\paragraph{Deep Learning-based Methods.}
These methods are mainly based on RNNs. GRU4REC~\cite{gru4rec} is the first to apply RNN in session-based recommendation where multiple GRU layers are stacked to make the prediction. NARM~\cite{narm} employs a hybrid encoder using attention mechanism to capture the user's main purpose in the current session. To alleviate the bias introduced by time series, STAMP~\cite{stamp} proposes a short-term attention/priority model with a novel attention mechanism to capture users' interest instead of using RNN.

\paragraph{GNN-based Methods.}
Due to the ability to model the complex relationships among items, GNNs have been received increasing attention in session-based recommendation. SR-GNN~\cite{sr-gnn} is the first to model sessions as graph structured data. It converts a session sequence into an unweighted directed graph and applies gated graph neural network (GGNN)~\cite{ggnn} to capture the complex item transitions in the session. FGNN~\cite{fgnn} formulates the session recommendation as a graph classification problem. It encodes a session sequence to a weighted directed graph and stacks multiple weighted graph attention layers to get accurate item representations.
GC-SAN~\cite{gc-san} also models the session sequence as a weighted directed graph. In GC-SAN, the local dependencies between items are captured by GGNN while the long-range dependencies are extracted through a self-attention network.
LESSR~\cite{lessr} converts a session into two kinds of graphs. One aims to generate a losslessly encoded graph and the other aims to address the long-range dependency problem. Two different layers are designed to learn item representations.
GCE-GNN~\cite{gce-gnn} learns two levels of item representations from both session graph and global graph, where the session graph is constructed based on the current session and the global graph is constructed from items' neighbors in all sessions. Graph attention network~\cite{gat} is applied to learn item representations. DAT-MDI~\cite{dat-mdi} combines dual tansfer with GNNs to learn cross-domain representation for session-based recommendation.

Unlike all the existing methods, we propose the first self-attention-based method that incorporates the transition relations between items for session-based recommendation.
\section{Preliminary: Self-Attention Mechanism}
The self-attention mechanism is originally used to model the direct relation between a source vector and any other context vectors (including itself) in a sequence. It first computes the attention scores between the source vector and the other context vectors. Then, all the context vectors are aggregated based on the attention scores to get the new representation of the source vector. Formally, given a source vector $\bm{x} \in \mathbb{R}^d$ and the set of context vectors $\{\bm{y}_1,\bm{y}_2,...,\bm{y}_n\}$ with the same dimension, the attention scores are calculated by their dot-product after applying two different linear transformations.
\begin{align}
    score(\bm{x}, \bm{y}_i) = \left(\bm{W}_q\bm{x}\right)^T\bm{W}_k\bm{y}_i
\end{align}
where $\bm{W}_q,\bm{W}_k \in \mathbb{R}^{d \times d'}$ are learnable parameter matrices.

Then the attention scores are normalized and a softmax function is applied to get the final attention scores.
\begin{align}
    a_i = \frac{exp\left(score\left(\bm{x}, \bm{y}_i\right)/\sqrt{d'}\right)}
    {\sum_{j=1}^{n}exp\left(score\left(\bm{x}, \bm{y}_j\right)/\sqrt{d'}\right)}
\end{align}

Finally, all the context vectors are aggregate based on the attention score to get the final output $attn$ (the new representation of the source vector) after a linear transformation.
\begin{align}
    attn = \sum_{i=1}^{n}a_i \bm{W}_v \bm{y}_i
\end{align}
where $\bm{W}_v \in \mathbb{R}^{d' \times d}$ is a learnable parameter matrix.


\section{Methodology}
\subsection{Problem Definition}
The session-based recommendation problem is to predict the next item based on the historically interacted items of the active session~\cite{survey}. We first present a formulation of session-based recommendation as below.

Let set $V=\{v_1, v_2,...,v_n\}$ denote all items that appear in all sessions. An anonymous session  can be represented as a sequence $S_s=\{v_1^s, v_2^s,...,v_l^s\}$, where each item corresponds to an item in $V$. $S_s$ indicates chronologically ordered user-item interactions in a continuous period of time. The goal of session-based recommendation is to predict the top-$K$ items from $V$ that are most likely to be clicked by the user in a session. A typical model for session-based recommendation outputs probabilities $\mathbf{\hat{\bm{y}}}$ for all items, where each element in $\mathbf{\hat{\bm{y}}}$ represents the recommendation score for the corresponding item. Then the top-$K$ items with highest scores will be recommended to the user. 

Figure \ref{fig:mode_architecture} shows the workflow of the proposed method TRASA for session-based recommendation, which consists of four stages: graph construction, item representation learning, session representation, and prediction.

\subsection{Graph Construction}
In our method, a session $S_s$ is modeled as a directed graph $\mathcal{G}_s=(\mathcal{V}_s, \mathcal{E}_s)$. $\mathcal{V}_s \subseteq V$ is the node set corresponding to unique item in the original session and $\mathcal{E}_s = \{e_{ij}^s\}$ is the edge set. Inspired by~\cite{gce-gnn}, $\mathcal{E}_s$ has four types. As shown in Figure \ref{fig:mode_architecture} (a), if $(v_i^s, v_j^s)$ are adjacent items in session $S_s$, two different types of directed edges are added between them. One is from $v_i^s$ to $v_j^s$, i.e., \textit{NXT} edge. The other is from $v_j^s$ to $v_i^s$, i.e., \textit{PRE} edge. In addition, if $(v_j^s, v_i^s)$ also appears in session $S_s$, the edge between $v_i^s$ and $v_j^s$ will be changed to \textit{NPL} edge. It is obvious that edges in the graph depict the click order of items in the original session. Finally, a self loop is added to each item because we assume that an item is also related to itself. We call this type \textit{SELF} edge. We assign each type of edge a learnable embedding, which is later used to encode the transition relations between items. 


\subsection{Item Representation Learning}
After constructing the graph, we encode the shortest path between any two items as their transition relation and use the self-attention mechanism to learn the item representations. 

\subsubsection{Relation Encoder}
Given the graph, we calculate the shortest path for each pair of items and use it to encode their transition relation~\cite{graph-transformer}. Suppose the shortest path between item $v_i^s$ and $v_j^s$ is $v_i^s \xrightarrow{NEXT} v_k^s \xrightarrow{NEXT} v_{k'}^s \xrightarrow{NPL} v_j^s$. Since the user's clicks are in chronological order, we use recurrent neural network with GRU to transform the shortest path to a fixed-length vector. Specifically, we employ the bi-directional GRU to encode the path. The input for the GRU is the edge sequence \textit{NEXT} $\rightarrow$ \textit{NEXT} $\rightarrow$ \textit{NPL} and the last hidden states of the GRU in both directions are concatenated as the final relation for item $v_i^s$ and $v_j^s$.

\subsubsection{Relation Aware Self-Attention}
Next, we calculate the attention score $a_{ij}$ between item $v_i^s$ and $v_j^s$. Their relation is $\bm{r}_{ij}$ which is encoded by the relation encoder. To incorporate transition relations in self-attention layers, $r_{ij}$ is first split into two relations $\bm{r}_{i \rightarrow j}$ and $\bm{r}_{j \rightarrow i}$, corresponding to the relation from $v_i^s$ to $v_j^s$ and the relation from $v_j^s$ to $v_i^s$ separately:
\begin{align}
    [\bm{r}_{i \rightarrow j};\bm{r}_{j \rightarrow i}] = \bm{W}_r \bm{r}_{ij}
\end{align}
where $\bm{W}_r \in \mathbb{R}^{2d \times d}$ is the learnable parameter and $d$ is the dimension of relation embeddings.

Then, the attention score is calculated based on both item embeddings and their transition relations:
\begin{align}
    a_{ij} = (\bm{h}_i + \bm{r}_{i \rightarrow j})\bm{W}_q^T \bm{W}_k(\bm{h}_j + \bm{r}_{j \rightarrow i})
\end{align}
Following the transformer~\cite{transformer} architecture, we use multi-head attention to get multi-head outputs and they are concatenated and projected to get the final values. As shown in Figure \ref{fig:mode_architecture} (b), a feed forward network and two residual connections are also applied in self-attention layers. Finally, mulitple self-attention layers are stacked to achieve better performance.




\subsection{Session Representation}
After generating the item representations $H_g = \{\bm{h}_1^g, \bm{h}_2^g,...,\bm{h}_m^g\}$. We need to get the final session representation. We first convert the graph to its original sequence $H_s = \{\bm{h}_1^s, \bm{h}_2^s,...,\bm{h}_l^s\}$.
Inspired by~\cite{gce-gnn}, we introduce reversed position embedding to keep the sequential position information. Intuitively, items have the same distance from the last item in the session should have the same importance for predicting the next click. The reversed position embeddings can portray this importance accurately. The position embeddings correspond to a learnable vector set $P_s = \{\bm{p}_1, \bm{p}_2,...,\bm{p}_l\}$, which are added to the item representations to get the final item representations.
\begin{align}
    \bm{h}_i^s = \bm{h}_i^s + \bm{p}_{l-i+1}
\end{align}



Since items clicked by users recently can reflect their current interests effectively, the recent clicked items have a strong influence to predict the next clicked item. Following~\cite{sr-gnn,lessr}, we regard the last item representation in a session as a user's current interest, i.e., $\bm{h}_l^s$. Then the user's preference is computed based on all the items in the session and the current interest. Since each item in a session should have different contributions to the final session representation, we aggregate all the item representations using different weights.
\begin{align}
    \bm{s_h} = \sum_{i=1}^{l} \gamma_i \bm{h}_i^s
\end{align}
where the weight $\gamma_i$ is decided by both the current item representation and the user's current interest (the last item representation). A soft-attention mechanism is applied to compute the weight.
\begin{align}
    \epsilon_i &= \bm{q}^T \left( \bm{W}_4 \bm{h}_i^s + \bm{W}_5 \bm{h}_l^s + \bm{b}_3 \right) \\
    \gamma_i &= \frac{exp\left(\epsilon_i\right)}{\sum_{j=1}^{l} exp\left(\epsilon_j\right)}
\end{align}
where $\bm{W}_5,\bm{W}_6 \in \mathbb{R}^{d \times d}$, $\bm{q} \in \mathbb{R}^d$, and $\bm{b}_3 \in \mathbb{R}^d$ are learnable parameters.

\subsection{Prediction}
After obtaining the final representation for each session, we can use it to compute the probabilities for all candidate items in $V$. To alleviate the long-tail problem in recommendation~\cite{niser} and make our model get a better convergence, we perform L2 normalization for all item embeddings in $V$. Then, for each item in $V$, we calculate its score based on its embedding and the session representation as follows:
\begin{align}
    \hat{\bm{z}}_i = \bm{s_h}^T \bm{h}_{v_i}
\end{align}
The softmax function is leveraged to get the final predicted probability:
\begin{align}
    \hat{\bm{y}} = softmax \left( \hat{\bm{z}}\right)
\end{align}
where $\hat{\bm{y}}$ denotes the probabilities of all items in $V$ to be clicked next in the current session.

Finally, we employ cross-entropy of the prediction and the ground truth as the objective function to train model parameters.
\begin{align}
    \mathcal{L}(\hat{\bm{y}}) = -\sum_{i=1}^{n}\bm{y}_i log(\hat{\bm{y}}_i) + (1-\bm{y}_i)log(1-\hat{\bm{y}}_i)
\end{align}
where $\bm{y}$ is the one-hot encoding vector of the ground truth item and $\hat{\bm{y}}_i$ is the probability of item $v_i$.

\section{Experiments}
We conduct extensive experiments to verify the effectiveness of the proposed method TRASA and mainly answer the following questions:
\begin{itemize}
    \item \textbf{RQ1:} Does TRASA achieve the state-of-the-art performance compared to the existing methods?
    \item \textbf{RQ2:} How does each component of TRASA affect the performance?
    \item \textbf{RQ3:} How do different hyper-parameter settings affect the model performance?
\end{itemize}

\subsection{Datasets and Preprocessing}
We use three publicly available real-world datasets, named \textit{Diginetica}\footnote{https://competitions.codalab.org/competitions/11161}, \textit{Gowalla}\footnote{https://snap.stanford.edu/data/loc-Gowalla.html}, \textit{Last.fm}\footnote{http://ocelma.net/MusicRecommendationDataset/lastfm-1K.html}. The three datasets are commonly used in literatures of session-based recommendation. 
We first preprocess these three datasets following~\cite{lessr}. Sessions of length 1 and items with less than 5 occurrences in all sessions are filtered in all three datasets. After that, we apply data augmentation for all sessions. For example, given a session $S=\{v_1, v_2,...v_l\}$, we generate the sequences and its corresponding labels as $([v_1], v_2), ([v_1, v_2], v_3),..., ([v_1, v_2,...v_{l-1}], v_l)$ for both training data and testing data. The statistics of the three preprecessed datasets are summarized in Table \ref{dataset_statistics}.

\begin{table}
\centering
\begin{tabular}{lrrr}
\toprule
Statistic           & Diginetica & Gowalla    & Last.fm \\
\midrule
No. of Clicks       & 981,620    & 1,122,788  & 3,835,706 \\
No. of Sessions     & 777,029    & 830,893    & 3,510,163 \\
No. of Items        & 42,596     & 29,510     & 38,615    \\
Average length      & 4.80       & 3.85       & 11.78     \\
\bottomrule
\end{tabular}
\caption{Statistics of datasets used in the experiments}
\label{dataset_statistics}
\end{table}

\subsection{Baseline Algorithms and Evaluation Metrices}
We compare TRASA with the following baselines that involves traditional methods, deep learning-based methods, and the SOTA GNN-based methods.

\begin{itemize}
    \item \textbf{Item-KNN}~\cite{Item-KNN} recommends items based on the consine similarity between items.
    \item \textbf{FPMC}~\cite{fpmc} utilizes both Markov chains and matirx factorization to capture user's interest.
    \item \textbf{GRU4Rec}~\cite{gru4rec} is a RNN-based method which stacks multiple GRU layers to model sessions.
    \item \textbf{NARM}~\cite{narm} employs a hybrid encoder using attention mechanism to capture the user's main purpose in the current session.
    \item \textbf{STAMP}~\cite{stamp} uses a short-term attention/priority model with attention mechanism to capture user's interest.
    \item \textbf{SR-GNN}~\cite{sr-gnn} converts a session to a graph and applies GNN to learn item representations.
    \item \textbf{GC-SAN}~\cite{gc-san} utilizes GGNN to capture the local dependencies and extracts the long-range dependencies by a self-attention network.
    \item \textbf{LESSR}~\cite{lessr} proposes to convert the session into two kinds of graphs and applies corresponding GNN layers to learn item embeddings. 
    \item \textbf{GCE-GNN}~\cite{gce-gnn} learns two levels of item representations from the session graph and global graph through graph attention networks and uses both of them to make prediction.
\end{itemize}

Note that all the methods use the same preprocessed datasets. The reason for no comparison with DAT-MDI~\cite{dat-mdi} is that DAT-MDI uses multiple datasets to implement cross-domain recommendation and no publicly available code has been found for DAT-MDI. Following previous works~\cite{sr-gnn,gce-gnn}, we adapt two commonly used metrics: \textbf{P@N} (Precision) and \textbf{MRR@N} (Mean Reciprocal Rank) as evaluation metrics.

\subsection{Parameter Setup}
In TRASA, we set the item embedding size $d=64$ for all three datasets. The batch-size is 512 for Diginetica and Gowalla and 2048 for Last.fm. All parameters are initialized using a Gaussian distribution with a mean of 0 and a standard deviation of 0.02. We use the Adam optimizer with the initial learning rate 0.01, which will decay by 0.1 after every 3 epoch. The L2 penalty is set $10^{-5}$ and dropout ratio is 0.2. We select other hyper-parameters using a validation set which is a random 10\% subset of training data.

\begin{table}
\centering
\renewcommand\arraystretch{1.3}
\resizebox{\linewidth}{!}{
    \begin{tabular}{ccc|cc|cc}
    \toprule
    \multirow{2}*[-3pt]{Methods} & \multicolumn{2}{c}{Diginetica} & \multicolumn{2}{c}{Gowalla} & \multicolumn{2}{c}{Last.fm} \\
    \cmidrule{2-7}
             &P@20 &MRR@20   &P@20 &MRR@20 &P@20 &MRR@20 \\
    \midrule
    FPMC     &28.50 &7.67  &29.91 &11.45 &12.86 &3.78 \\
    Item-KNN &39.51 &11.22 &38.60 &16.66 &14.90 &4.04 \\
    GRU4Rec  &29.45 &8.22 &41.98 &18.37 &17.90 &5.39 \\
    NARM     &49.80 &16.57 &50.07 &23.92 &21.83 &7.59 \\
    STAMP    &45.64 &15.13 &50.18 &24.06 &22.01 &7.98 \\
    SR-GNN   &50.81 &17.31 &50.32 &24.25 &22.33 &8.23 \\
    GC-SAN   &50.90 &17.63 &50.68 &24.67 &22.64 &8.42 \\
    LESSR    &52.69 &18.29 &51.82 &\underline{25.88} &23.39 &\underline{9.04} \\
    GCE-GNN  &\underline{54.71} &\underline{19.26} &\underline{53.56} &24.78 &\underline{23.91} &8.33 \\
    \midrule
    \textbf{TRASA}     &\textbf{55.15} &\textbf{19.47} &\textbf{54.22} &\textbf{26.20} &\textbf{24.60} &\textbf{9.35} \\
    \bottomrule
    \end{tabular}
}
\caption{Overall performance on three datasets (Bold is the best, the underline is the second best)}
\label{overall_comparison}
\end{table}

\subsection{Overall Comparison(RQ1)}
The performance comparison is summarized in Table \ref{overall_comparison}. It can be seen that TRASA can achieve the best performance across all three datasets in terms of the two evaluation metrics, which demonstrates the effectiveness of TRASA. 

For traditional methods, FPMC performs poorly, indicating that the assumption on the independence of successive items in Markov Chain-based methods is not sufficient. Item-KNN performs better than FPMC, which shows that considering the similarity between items does make a contribution to prediction. However, traditional methods are not competitive enough compared to other methods.

For deep learning-based methods, GRU4Rec performs worst. It simply stacks multiple GRU layers to make prediction. However, its performance improvement is still demonstrated on two datasets compared with traditional methods, which indicates that the deep learning-based methods have more powerful representation capabilities. NARM performs better than GRU4Rec. 
NARM combines both the sequential behavior and the main purpose to constitute users' preferences. It proves that simply applying RNN in session-based recommendation is insufficient. STAMP gets the best results in all deep learning-based methods by explicitly taking users' general and current interests into account. 

As the first GNN-based method, SR-GNN outperforms all deep learning-based methods. 
To capture the long-range dependencies, GC-SAN applies self-attention layers after the GNN layers and LESSR builds a short-cut graph to connect all items directly. Both GC-SAN and LESSR can achieve better performance than SR-GNN, proving that the long-range dependency can indeed affect the performance. GCE-GNN is the best among all the GNN-based methods, which indicates that considering two levels of item embeddings from session graph and global graph are beneficial for prediction. 

In contrast, TRASA outperforms the other methods consistently on all three datasets. It is because that we use self-attention mechanism to capture the long-range dependencies. Furthermore, to keep the item transition relations in the session, we employ the output of the relation encoder as supplementary information when computing the attention scores. 


\subsection{Deeper Model Analysis (RQ2)}
We conduct further analysis of TRASA to find out what exactly contributes to the performance improvement.
\paragraph{Impact of each component.} We conduct experiments to find out the effect of each component in TRASA.
\begin{itemize}
    \item WO-POS: without  the position embedding.
    \item WO-REL-POS: without the relation encoder and position embedding.
    \item WO-SAN: without  the self-attention layers.
\end{itemize}
From Table \ref{effects_components}, we can observe that self-attention layers play a very important role in TRASA. The model performance drops dramatically after removing self-attention layers, showing that it is necessary to learn the connections between different items.
By comparing WO-POS and WO-REL-POS, we can conclude that encoding the transition relations explicitly is very effective in session-based recommendation. In addition, the position embedding also makes a contribution to improve the model performance and utilizing the relation encoder and position embedding simultaneously can achieve a better model performance.   

\begin{table}[]
    \centering
    \renewcommand\arraystretch{1.2}
    \resizebox{\linewidth}{!}{
        \begin{tabular}{ccc|cc|cc}
        \toprule
        \multirow{2}*[-3pt]{Methods} & \multicolumn{2}{c}{Diginetica} & \multicolumn{2}{c}{Gowalla} & \multicolumn{2}{c}{Last.fm} \\
        \cmidrule{2-7}
        &P@20 &MRR@20 &P@20 &MRR@20 &P@20 &MRR@20 \\
        \midrule
        WO-POS        &55.07 &19.33 &53.97 &25.44 &24.24 &8.10\\
        WO-REL-POS   &51.59 &17.83 &52.25 &24.69 &23.47 &7.59\\
        WO-SAN         &34.36 &11.55 &32.59 &14.79 &16.44 &5.81\\
        \midrule
        \textbf{TRASA}  &\textbf{55.15} &\textbf{19.47} &\textbf{54.22} &\textbf{26.20} &\textbf{24.60} &\textbf{9.35}\\
        \bottomrule
        \end{tabular}}
    \caption{Effects of different components.}
    \label{effects_components}
\end{table}

\paragraph{Impact of different methods to represent sessions.} We explore different session representation methods.
\begin{itemize}
    \item SAN: directly using the self-attention function to get a session representation.
    \item SUM: using the sum of item embeddings as a session representation.
    \item GRAPH: getting a session representation without converting the graph to its original sequence. 
\end{itemize}
As shown in Table \ref{dif_ses_rep}, in the Diginetica dataset, SAN performs poorly compared to other methods. GRAPH has the closest performance to our method and SUM has a comparable performance as well. In the Gowalla and Last.fm datasets, both SAN and SUM have a comparable performance to our method and they perform better than GRAPH. In addition, we find that if we do not revert the graph to its original sequence, MRR@20 will drop. This phenomenon is consistent across all three datasets. To sum up, different session representation methods have different performance on different datasets but our method always has the best performance which demonstrates that aggregating the item embeddings based on user's current interest is an effective way to represent the session.

\begin{table}[]
    \centering
    \renewcommand\arraystretch{1.1}
    \resizebox{\linewidth}{!}{
        \begin{tabular}{ccc|cc|cc}
        \toprule
        \multirow{2}*[-3pt]{Methods} & \multicolumn{2}{c}{Diginetica} & \multicolumn{2}{c}{Gowalla} & \multicolumn{2}{c}{Last.fm} \\
        \cmidrule{2-7}
        &P@20 &MRR@20 &P@20 &MRR@20 &P@20 &MRR@20 \\
        \midrule
        SAN &52.45 &18.06 &54.05 &25.60 &24.55 &9.29 \\
        SUM &54.49 &19.11 &54.14 &25.69 &24.59 &9.23 \\
        GRAPH &55.09 &19.14 &53.50 &25.84 &24.58 &9.11 \\
        \midrule
        \textbf{TRASA} &\textbf{55.15} &\textbf{19.47} &\textbf{54.22} &\textbf{26.20} &\textbf{24.60} &\textbf{9.35} \\
        \bottomrule
        \end{tabular}}
    \caption{Performance comparison between different session representation methods.}
    \label{dif_ses_rep}
\end{table}

\subsection{Hyper-parameter study (RQ3)}
\begin{figure}
    \centering
    \subfloat[Diginetica]{
        \includegraphics[width=0.49\linewidth]{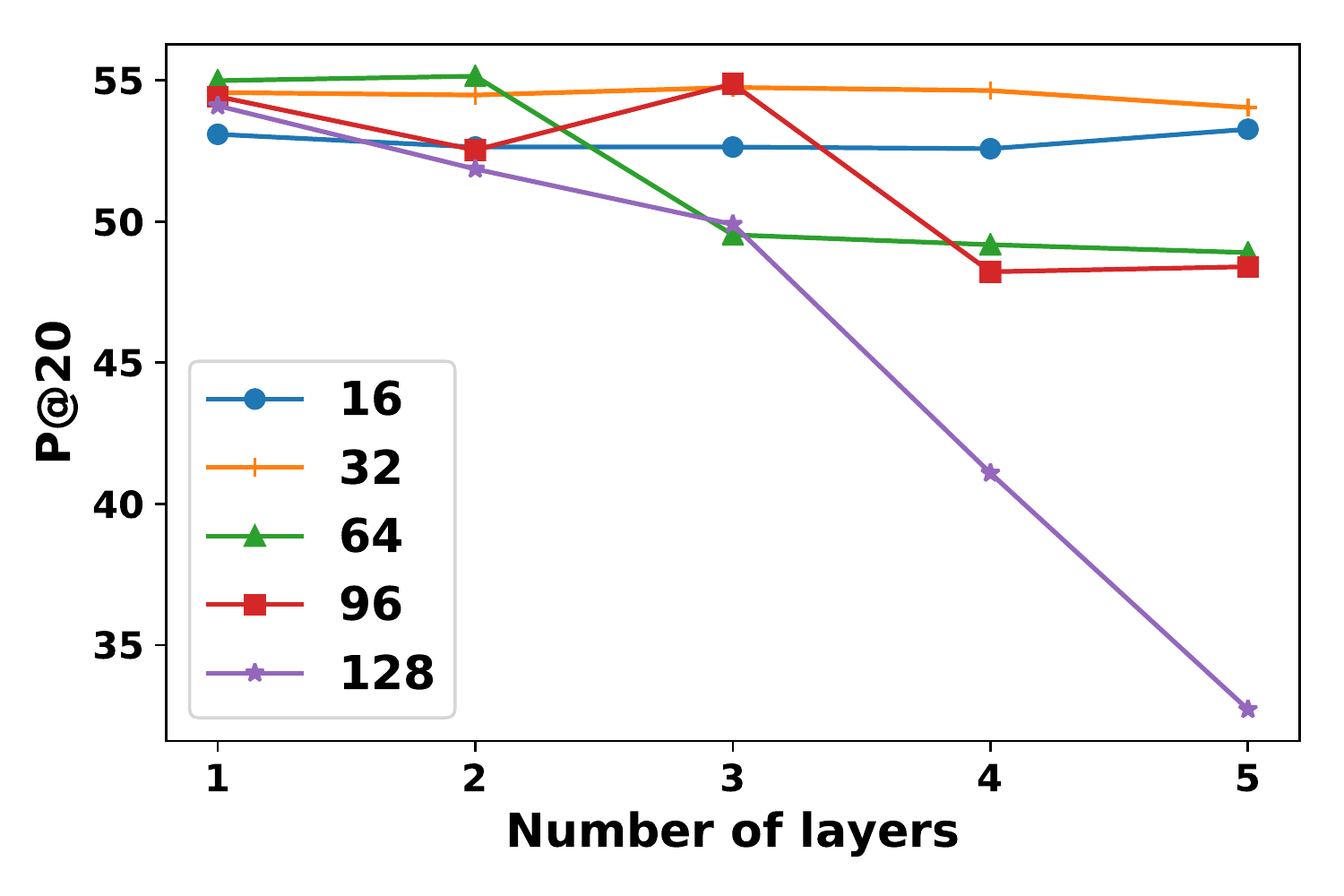}
        \label{fig:Caption_a}
    }
    \subfloat[Gowalla]{
        \includegraphics[width=0.49\linewidth]{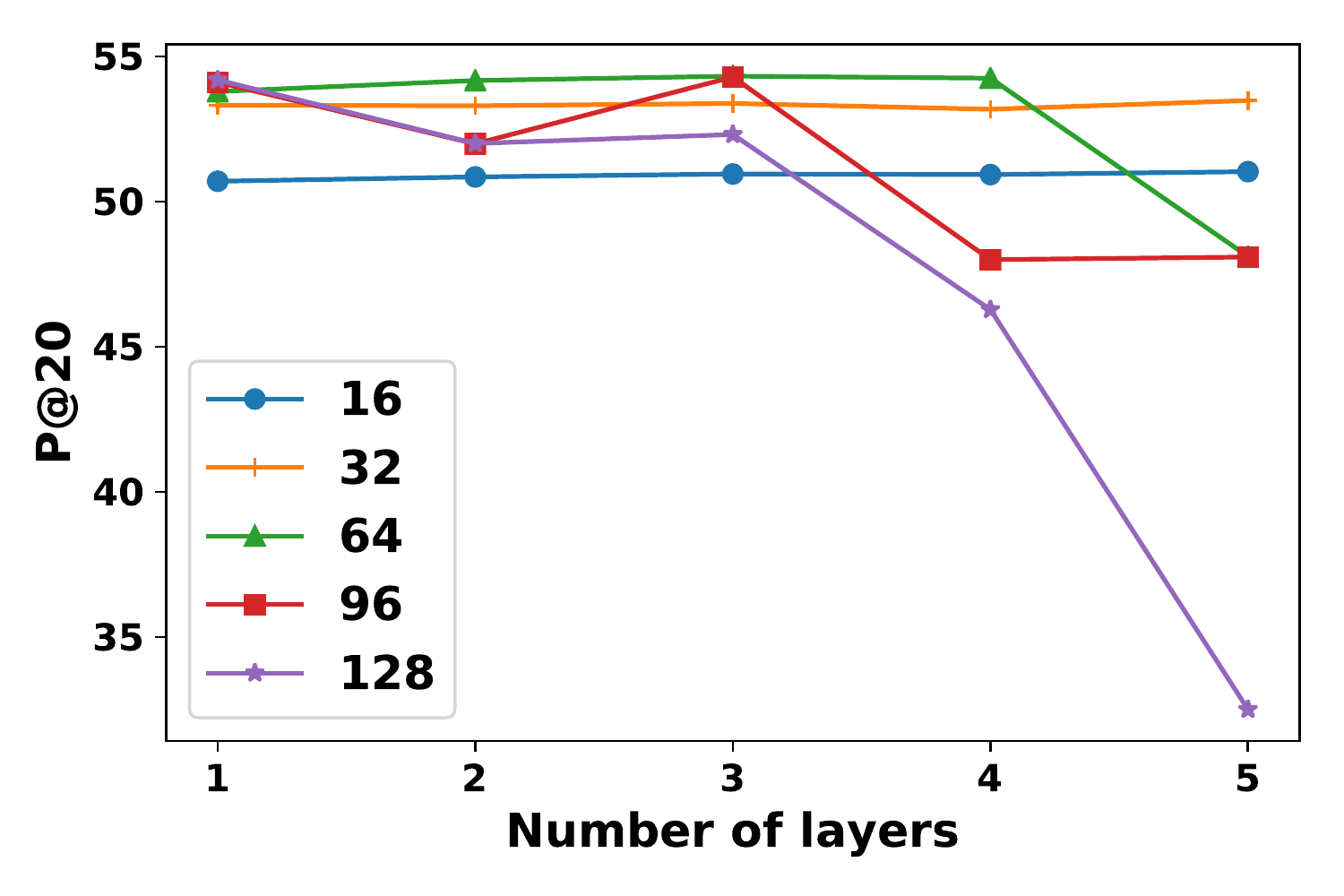}
        \label{fig:Caption_b}
    }
    \caption{Effects of the embedding size and the number of self-attention layers}
    \label{hp-study}
    \vspace{-2ex}
\end{figure}

In this section, we explore the effect of key hyperparameters in TRASA. We use the Diginetica and Gowalla datasets to study the effect of the embedding size and number of self-attention layers. The results are shown in Figure \ref{hp-study}.

\paragraph{Impact of the embedding size.} When the embedding size is small, TRASA does not perform well because embeddings with a small dimension cannot adequately encode all item characteristics. As the embedding size becomes larger, the performance improves accordingly. But when the embedding size gets too large (over 64 in our experimental setup), the performance drops, which indicates that larger embedding size is not always better due to the overfitting problem. 

\paragraph{Impact of the number of self-attention layers.} The models with different embedding sizes perform differently as the number of self-attention layers increases. When the embedding size is small, the model performance will keep growing as the number of layers increases. When the embedding size becomes larger, the model performance first increases and then decreases. In addition, with a larger embedding size, the model performance drops more and faster. It is probably because with a larger embedding size, the model representation capability is more powerful, and it becomes more prone to overfitting.


\section{Conclusion}
In this paper, we proposed a novel method called TRASA for session-based recommendation.
To the best of our knowledge, TRASA is the first to introduce the self-attention mechanism for item representation learning. Specifically, TRASA first converts the session into a directed graph. To model the item transition relations explicitly, TRASA encodes the shortest path between items as their transition relations. To capture the long-range dependencies, TRASA utilizes self-attention to make direct connections between any two items. Meanwhile, the transition relations are incorporated when computing the attention scores. Extensive experiments on three real-world datasets demonstrate that TRASA outperforms the existing state-of-the-art methods.


\bibliographystyle{named}
\bibliography{references}
\end{document}